\documentclass[aps,prb,floatfix,twocolumn,superscriptaddress]{revtex4-2}
\usepackage{graphicx}
\usepackage{epsfig}
\usepackage{psfrag}
\usepackage[usenames]{color}
\usepackage{xcolor}
\usepackage{hyperref}
\usepackage{dcolumn}
\usepackage{bm}
\usepackage{amsmath,amssymb,stmaryrd} 
\usepackage{bbm}

\hypersetup{colorlinks=true,citecolor={blue},linkcolor={blue},urlcolor={blue}}

\newcommand\id{\ensuremath{\mathbbm{1}}}

\makeatletter
\newcommand*{\rom}[1]{\expandafter\@slowromancap\romannumeral #1@}
\makeatother

\begin{document}

\title{Optical valleytronics of impurity states in two-dimensional Dirac materials}


\author{Dogyun Ko}
\affiliation{Center for Theoretical Physics of Complex Systems, Institute for Basic Science (IBS), Daejeon 34126, Korea}
\affiliation{Basic Science Program, Korea University of Science and Technology (UST), Daejeon 34113, Korea}

\author{A. V. Morozov}
\affiliation{Rzhanov Institute of Semiconductor Physics, Siberian Branch, Russian Academy of Sciences, Novosibirsk, 630090 Russia}

\author{V.~M.~Kovalev}
\affiliation{Rzhanov Institute of Semiconductor Physics, Siberian Branch, Russian Academy of Sciences, Novosibirsk, 630090 Russia}

\author{I.~G.~Savenko}
\affiliation{Center for Theoretical Physics of Complex Systems, Institute for Basic Science (IBS), Daejeon 34126, Korea}
\affiliation{Basic Science Program, Korea University of Science and Technology (UST), Daejeon 34113, Korea}

\date{\today}

\begin{abstract}
We analyze the valley selection rules for optical transitions from impurity states to the conduction band in two-dimensional Dirac materials, taking a monolayer of MoS$_2$ as an example.
We employ the analytical model of a shallow impurity potential which localizes electrons described by a
spinor wave function, and, first, find the system eigenstates taking into account the presence of two valleys in the Brillouin zone. 
Then, we find the spectrum of the absorbance and calculate the photon-drag electric current due to the impurity-band transitions, drawing the general conclusions regarding the valley optical selection rules for the impurity-band optical transitions in gapped Dirac materials.
\end{abstract}
\maketitle

The key idea of valleytronics is in using the valley index as an additional active degree of freedom of charge carriers~\cite{PhysRevLett.99.236809,PhysRevLett.108.196802} in gapped graphene~\cite{PhysRevLett.99.236809}, monolayers of transition metal dichalcogenides (TMDs)~\cite{PhysRevLett.108.196802}, among other two-dimensional (2D) Dirac materials. 
One of the representatives of TMDs is MoS$_2$: a material with a structure composed of molybdenum atoms sandwiched between pairs of sulfur atoms. 
In contrast to graphene, it is characterized by the inversion symmetry breaking and it possesses a large band gap with the width in the optical range, absent in monolayer graphene~\cite{Pan2014}. 
It represents a direct band gap material with the minima of the conduction band and maxima of the valence band located at points $K$ and $K'$ in reciprocal space.
Moreover, electrons in MoS$_2$ are subject to strong spin-orbital interaction, which also makes it different from graphene, where the spin-orbital interaction is relatively weak. 
This latter property, which is due to the electrons occupying d-orbitals in MoS$_2$, results in an extra band splitting~\cite{Silva_Guill_n_2016}.

It has been shown that the interband (between the conduction and valence bands) transitions in Dirac materials are valley-sensitive: at a given circular polarization of the external electromagnetic perturbation, the interband transitions occur predominantly in one valley since the electrons in each valley couple with a specific polarization of light~\cite{Kovalev_2018}. %
Switching to the opposite circular polarization changes the valley where the interband transitions take place~\cite{Zeng_2012}.
These optical selection rules are fulfilled exactly for interband optical transitions, where the electron momentum is a good quantum number. 

However, each material is to some extent disordered: it contains impurities, some of which are unintentional and emerge due to the imperfections of the growth technique, whereas some of the impurities are embedded intentionally in order to enhance electronic (or other) properties of the sample.
As a result, there emerge additional donor and acceptor energy levels in the vicinity of the conduction and valence bands, respectively.
Then, if the sample is exposed to external light
with the frequency smaller than the band gap of the material, the optical properties become mostly determined by the electron transitions from donor impurities to the conduction band and from the valence band to the acceptor states~\cite{Li1993}. 
In this case, the electron states on impurities are characterized by some quantum numbers instead of the translational momentum due to localization. 
The theoretical description of optical transitions from these states to the bands and the analysis of the corresponding optical selection rules, which take into account the valley quantum number, represents an important problem of valley optoelectronics, and it has been studied theoretically. 
In particular,  work~\cite{PhysRevLett.121.167402} presents a numerical study of a special type of disorder: the vacancy defects.
%
%
%
\begin{figure}[t!]
\centering
\includegraphics[width=0.48\textwidth]{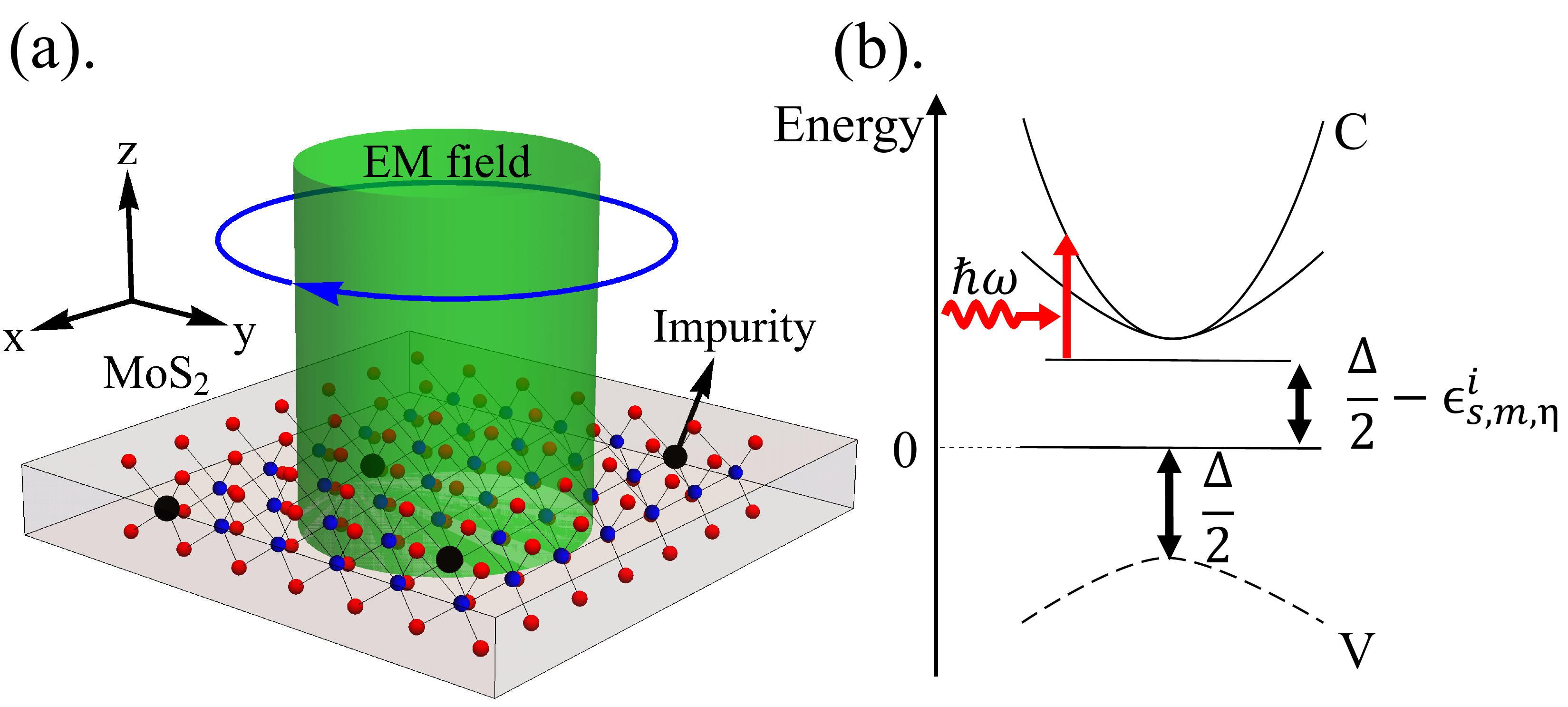}
\caption{System schematic. (a) A monolayer of MoS$_2$ (with impurities) exposed to a circularly polarized electromagnetic (EM) field of light (green cylinder). 
(b) The bandstructure of MoS$_2$. The band gap is $\Delta$, and the system is exposed to an EM field with frequency $\omega$.
}
\label{Fig1}
\end{figure}
%
%
%

Usually, the energy gap between the impurity states and the conduction band corresponds to the terahertz (THz) frequency range (10-20 meV). 
It can be used to design (pulsed) terahertz radiation detectors. 
In such a detector, a polarized optical signal is transformed to electric current (to be measured).
The analysis of the optical selection rules here is thus of utmost importance for applications.
We want also to mention another potential application of the theory of impurity-band transitions.
It has been recently been proposed that TMD monolayers containing atomic impurities can be used as single-photon emitters~\cite{barthelmi2020atomistic}.
Utilizing artificially-created atomic vacancies, one can achieve the single-photon regime of operation.
This is one of fast-developing topics of research recently. 

For the fundamental purposes and to design the devices, it is useful to study the general properties of defects of any type, and an analytical analysis would be beneficial here.
However, one of the problems to face is that the simple model, which assumes the impurity potential energy to be the Dirac delta-function~\cite{doi:10.1063/1.3556738}, is not applicable in the case of a Dirac Hamiltonian since the electron wave function becomes singular exactly at the center of coordinates~\cite{PhysRevLett.96.126402}.

In this Letter, we build a theory of impurity-band transitions in 2D Dirac materials, utilizing and modifying the model of zero-radius impurity potential, which is frequently used for the description of shallow impurities in semiconductors and semiconductor nanostructures~\cite{PhysRevLett.96.126402, pakhomov1996local}. 
We investigate the optical properties of disordered TMDs, examining the light absorption and photoinduced transport effects, accounting for the spin-orbital coupling of electrons.
%
We study the behavior of drag electric current density and the absorption coefficient for different key parameters of the sample and different polarizations of the incident light.
It should be mentioned, that the generation of the electric current in 2D Dirac materials due to the \textit{interband} optical transitions (the photon and phonon drag effects) has been extensively studied~\cite{PhysRevB.81.165441, golub2011valley, PhysRevLett.122.256801, PhysRevB.102.235405, PhysRevB.102.045407}, but the impurity-band transitions have not been addressed.


\textit{Hamiltonian and eigenstates. }
Light absorption is governed by microscopic transitions of electrons from the
bound impurity states to the conduction band. 
The Hamiltonian of the electron bound at the attractive potential $u(\textbf{r})$ reads
\begin{equation}
\label{EqHam1}
H=\left(\frac{\Delta}{2}\sigma_z + \textbf{v}\cdot\textbf{p}\right)\otimes \id - \frac{\lambda\eta}{2}(\sigma_z-\id)\otimes\hat{s}_z+u(\textbf{r}),
\end{equation}
where $\Delta$ is a band gap, $\textbf{v} = v(\eta\sigma_x,\sigma_y)$ is the velocity operator,
$\textbf{p}$ is the electron momentum, and $\sigma_\alpha$ with $\alpha=x,y,z$ the Pauli matrices of pseudospin. 
The index $\eta=\pm1$ indicates the valley; 
$\lambda$ is intrinsic spin-orbital coupling; $\hat{s}_z$ is the matrix of the electron spin.
The first term in Hamiltonian~(\ref{EqHam1}) 
describes a two-band model of gapped graphene (or a band structure of a TMD material). 
We consider a shallow impurity potential, $u(\textbf{r})$, thus we assume that the ionization potential of the donor is much smaller than $\Delta$.

To find the eigenfunctions and eigenenergies, we write the Schr\"odinger equation in the momentum representation,
\begin{eqnarray}
\label{EqMatHam}
&&
\begin{pmatrix}
\frac{\Delta}{2}-E & vp_{-} 
\\
vp_{+} & -\frac{\Delta}{2}+s\lambda\eta-E 
\end{pmatrix}
\chi_{s,m}(\textbf{p})
\\
\nonumber
&&~~~~~~~~~~~~~~~~~~~~
+\int\frac{d\textbf{p}'
}{(2\pi\hbar)^2}
u({\textbf{p}-\textbf{p}'})\chi_{s,m}(\textbf{p}')
=0,
\end{eqnarray}
where $s=\pm 1$,  $p_\pm=\eta p_x\pm ip_y=pe^{\pm i\eta\varphi}$ with $\varphi$ the angle of the vector $\textbf{p}$ with respect to $x$-axis, $m$ is the eigenvalue of the z-projection of the electron angular momentum (the quantum number which characterizes electron localized on impurity),
\begin{eqnarray}
&&u(\textbf{p}-\textbf{p}')
=
2\pi\int^\infty_0rdru(r)J_0(|\textbf{p}-\textbf{p}'|r)
\\
\nonumber
&&~~~~~=
2\pi\sum_{k}\int^\infty_0rdru(r)J_k(pr)J_k(p'r)\cos k(\varphi-\varphi '),
\end{eqnarray}
where 
$J_k(x)$ are the $k$-order Bessel functions. 
We search for the spinor eigenfunctions in the form,
\begin{equation}
\label{EqEig1}
\chi_{s,m}(\textbf{p})=\left(
  \begin{array}{c}
    a_{s,m}(p)e^{i m\varphi}\\
    b_{s,m}(p)e^{i(m+\eta)\varphi}
  \end{array}
\right)
\end{equation}
since this form reflects the axial symmetry, 
and the eigenstates are characterized by the angular momentum projection with quantum number $m$.
Substituting Eq.~\eqref{EqEig1} in Eq.~\eqref{EqMatHam} and performing the integration over $\varphi'$, we find the system of equations for the coefficients $a_{s,m}$ and $b_{s,m}$,
\begin{eqnarray}
\label{Bessel}
&&0=
\begin{pmatrix}
\frac{\Delta}{2}-E & vp
\\
vp & -\frac{\Delta}{2}+s\lambda\eta-E
\end{pmatrix}
\begin{pmatrix}
a_{s,m}(p)\\
b_{s,m}(p)
\end{pmatrix}
\\
\nonumber
&&+
\int\limits^\infty_0rdru(r)\int\limits^\infty_0
\frac{p'dp'}{\hbar^2}
\begin{pmatrix}
J_m(pr)J_m(p'r)a_{s,m}(p')\\
J_{(m+\eta)}(pr)J_{(m+\eta)}(p'r)b_{s,m}(p')
\end{pmatrix}.
\end{eqnarray}
%
To draw principal conclusions, we can now simplify these equations.
For a shallow impurity ($\epsilon^i_{s,m,\eta}\ll\Delta$) and low enough temperatures only the low-lying impurity states are occupied. Then, we can consider the transitions from impurity states corresponding to $m=0$ and $m=\pm1$ levels only.
Assuming that the potential of each impurity $u(r)$ is sharply peaked in the vicinity of its center $r=0$ and it rapidly decreases with $r$~\cite{[{On one hand, in the framework of our model we assume that the potential is narrow enough, thus, we make certain qualitative simplifications in Eq.~\eqref{simple}. On the other hand, in order to disregard the intervalley mixing (and thus, disregard the splitting of the impurity states due to the intervalley mixing of electron wave functions on impurities), the potential has to be wide enough. Formally, if we denote as $a$ the characteristic size of the impurity potential and as $L$ the characteristic width of the wave function of electron localized on the impurity, one should fulfill $L\gg a\gg 1/(2K_0)$, where $2K_0$ is the distance between nonequivalent valleys in reciprocal space}]c2}, we can take the Bessel functions under the integral $r=0$. For the $m=0$ state, $J_0(0)=1$ and $J_{\eta=\pm1}(0)=0,$ and we find the simplified form of Eq.~\eqref{Bessel},
\begin{equation}
\label{simple}
\begin{split}
\begin{pmatrix}
\epsilon^i_{s,0,\eta} & vp \\
vp & -\Delta+s\lambda\eta
\end{pmatrix}
\begin{pmatrix}
a_{s,0} \\ b_{s,0}
\end{pmatrix}
+
\begin{pmatrix}
Au_0 \\ 0
\end{pmatrix}
=0,
\end{split}    
\end{equation}
where 
\begin{equation}
\label{Parms}
A=\int^\infty_0\frac{p'dp'}{\hbar^2}a_{s,0}(p');~~~ u_0 = \int^\infty_0u(r)rdr.
\end{equation}
The solution of ~Eq .\eqref{simple} reads 
\begin{equation}\label{WF}
\begin{pmatrix}
a_{s,0} \\ b_{s,0}
\end{pmatrix}
=-
\hbar v
\sqrt{\frac{2\pi\epsilon^i_{s,0,\eta}}{\Delta}}
\begin{pmatrix}
\frac{\Delta-s\lambda\eta}{(vp)^2+\epsilon^i_{s,0,\eta}(\Delta-s\lambda\eta)}
\\
\frac{vp}{(vp)^2+\epsilon^i_{s,0,\eta}(\Delta-s\lambda\eta)}
\end{pmatrix}.
\end{equation}
%

For the $m = 1, \eta = -1$ state, 
$J_0(0)=1$, $J_1(0)=0$, and $J_2(0)=0$,  Eq.~\eqref{Bessel} can be simplified,
%
\begin{equation}\label{M1eq}
\begin{split}
\begin{pmatrix}
\epsilon^i_{s,1,-1} & vp \\
vp & -\Delta-s\lambda
\end{pmatrix}
\begin{pmatrix}
a_{s,1} \\ b_{s,1}
\end{pmatrix}
+
\begin{pmatrix}
0 \\ Bu_0
\end{pmatrix}
=0,
\end{split}    
\end{equation}
where now
\begin{equation}
\label{defB}
B=\int^\infty_0\frac{p'dp'}{\hbar^2}b_{s,1}(p').
\end{equation}
The solution of ~Eq. \eqref{M1eq} reads
\begin{equation}
\begin{pmatrix}
a_{s,1} \\ b_{s,1}
\end{pmatrix}
=-
\hbar v
\sqrt
{
\frac
{2\pi(\Delta^2-\lambda^2)}
{\Delta\epsilon^i_{s,1,-1}}
}
\begin{pmatrix}
\frac{vp}{(vp)^2+\epsilon^i_{s,1,-1}(\Delta+s\lambda)}
\\
\frac{-\epsilon^i_{s,1,-1}}{(vp)^2+\epsilon^i_{s,1,-1}(\Delta+s\lambda)}
\end{pmatrix}.
\end{equation}
The energy $\epsilon^i_{s,1,-1}$ of this state can be also found using the definition Eq.~\eqref{defB}. We see that within the framework of the shallow-impurity model, the state $m=1$ forms `under' the $\eta=-1$ valley (and vise versa).
In other words, the following rule holds: $m+\eta=0$ for $m=\pm1$ states.

The electron states in the
conduction band are described by the wave function,
%
\begin{equation}
\psi_{s,\eta}(\textbf{p})=\left(
  \begin{array}{c}
   \cos\left(\frac{\theta_{s,\eta}}{2}\right) \\
   \sin\left(\frac{\theta_{s,\eta}}{2}\right)e^{i\eta\varphi_{\textbf{p}}}
  \end{array}
\right),
\end{equation}
where we use the notations $\cos\theta_{s,\eta}=(\Delta-s\lambda\eta)/2E_{s,\eta}(\textbf{p})$ and $\sin\theta_{s,\eta}=\eta vp/E_{s,\eta}(\textbf{p})$ 
with the conduction band electron energy $E_c(\textbf{p}) = s\eta\lambda/2 + E_{s,\eta}(\textbf{p})$, $E_{s,\eta}(\textbf{p})=\sqrt{(vp)^2+\left[(\Delta-s\lambda\eta)/2\right]^2}$.
Since the transitions from the impurity state with a given valley number $\eta$ to the conduction band of the other valley $\eta'\neq\eta$ are strongly suppressed due to the large distance between the valleys in the reciprocal space~\cite{Gorycaeaau4899}, the main contribution to the light absorption comes from the impurity-band transitions with the same valley number $\eta' = \eta$. 
From the point of view of applications, the most interesting is the circularly-polarized EM field case. 
The Hamiltonian describing the interaction of electrons with the external EM perturbation reads $\hat{V}(\textbf{}{r},t)=-e\textbf{v}\cdot\mathbf{A}(\textbf{r},t)$,
%
%
where $\mathbf{A}(\mathbf{r},t) = \mathbf{A}_0 exp(i\mathbf{kr}-i\omega t) + \mathbf{A}^{*}_0 exp(-i\mathbf{kr}+i\omega t)$ is the vector potential of EM field.
Here, $\textbf{A}_0 = A_0\hat{x}+A_0i\sigma\hat{y}$ with $\sigma$ the light polarization, $\hat{x}$ and $\hat{y}$ the unity vectors in the corresponding directions in direct space, $\mathbf{k}$ and $\omega$ being the photon wave vector and frequency, respectively.
\begin{figure}[t!]
\centering
\includegraphics[width=0.4\textwidth]{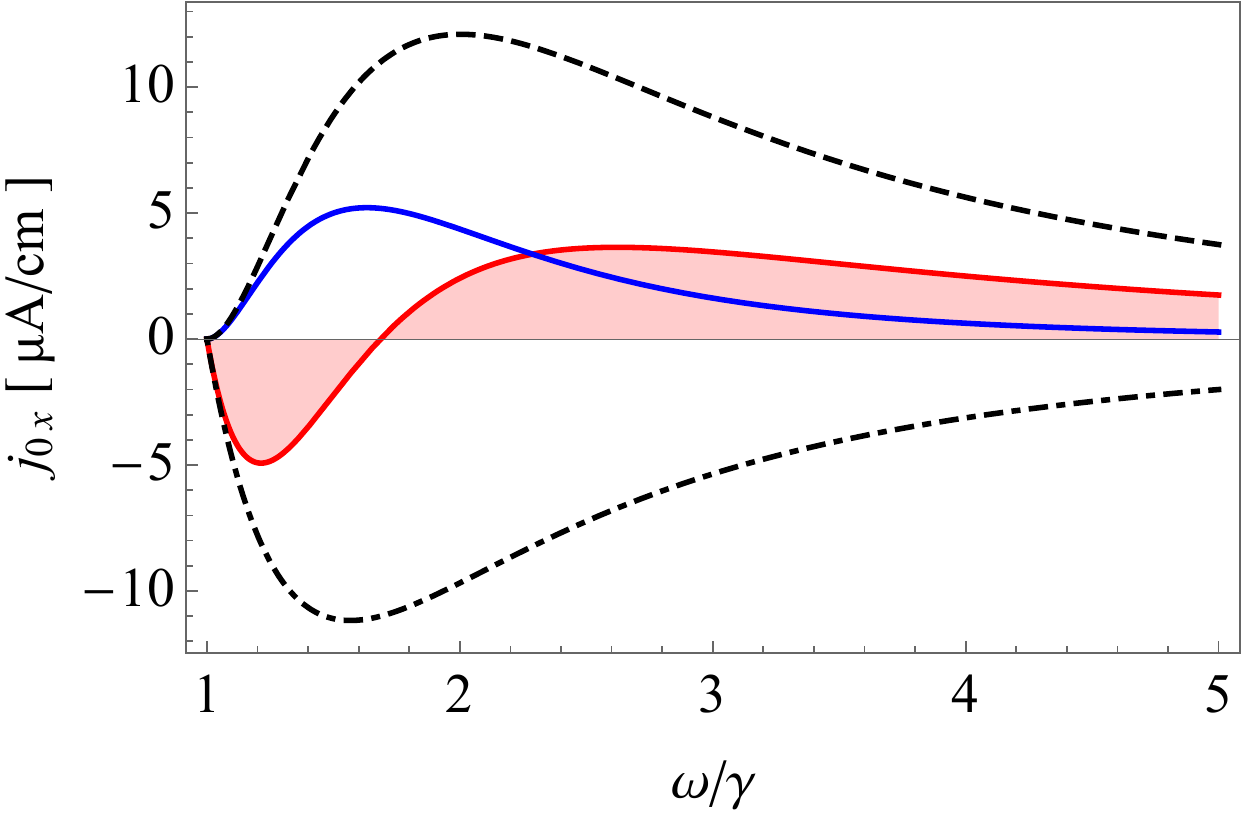}
\caption{Spectrum of electric current density due to the transitions from $m=0$, $\eta=-1$ 
impurity states to spin-up conduction band states for the polarization of light $\sigma = -1$ (red) and $\sigma = 1$ (blue); $\gamma=\epsilon^{i}_{1,0,-1}/\hbar$, 
where $\epsilon^{i}_{1,0,-1} = 10$~meV is the energy of impurity counted from the bottom of the conduction band.
Black curves show the positive and negative contributions to current in $\sigma = -1$ case.
We used the density of impurities
$n_i \approx 5\times10^{12}$~cm$^{-2}$; 
electron relaxation time $\tau=2\times10^{-13}$~s, velocity $v=at/\hbar$, the lattice constant of MoS$_{2}$ $a=3.193$~\AA,
effective hopping integral $t=1.10$~eV~\cite{Hatami_2014}, amplitude of light
$A_0=3.8\times10^{12}$ eV$\cdot$ s/C$\cdot m$, the band gap $\Delta=1.16$~eV, and the spin-orbit coupling strength $\lambda=75$~meV.
}
\label{Fig2}
\end{figure}
%


\textit{Electric current density. }
The general expression for the (partial) component of photon-drag electric current density, corresponding to electron transitions from the impurity state with a quantum number $m$ to the conduction band, reads $(\alpha = x, y)$
\begin{equation}
\label{EqCurrentMain}
j_{m\alpha}=\frac{2\pi en_i\tau}{\hbar}\int
\frac{v_\alpha(\textbf{p})d\textbf{p}}{(2\pi\hbar)^2}
|M_m(\textbf{p},\textbf{k})|^2
\delta(E_{c}(\textbf{p})-E_i-\hbar \omega),
\end{equation}
where $n_i$ is the impurity concentration, $e$ is the elementary charge, $\tau$ is the electron relaxation time in conduction band, $M_m(\mathbf{p},\mathbf{k})=\langle\psi_{s,\eta}(\mathbf{p})|\hat{V}|\chi_{s,m}(\mathbf{p}-\mathbf{k})\rangle$ is the matrix element of impurity-band transitions, and the electron velocity components read $v_x(\textbf{p})=\eta\sin\theta_\textbf{p}\cos\varphi_\textbf{p}$ and $v_y(\textbf{p})=-i\eta\sin\theta_\textbf{p}\sin\varphi_\textbf{p}$; 
$E_i = \Delta/2 - \epsilon^i_{s,m,\eta}$ is the impurity energy level.
%
%
%
%
%
%
%
\begin{figure}[t!]
\centering
\includegraphics[width=0.4\textwidth]{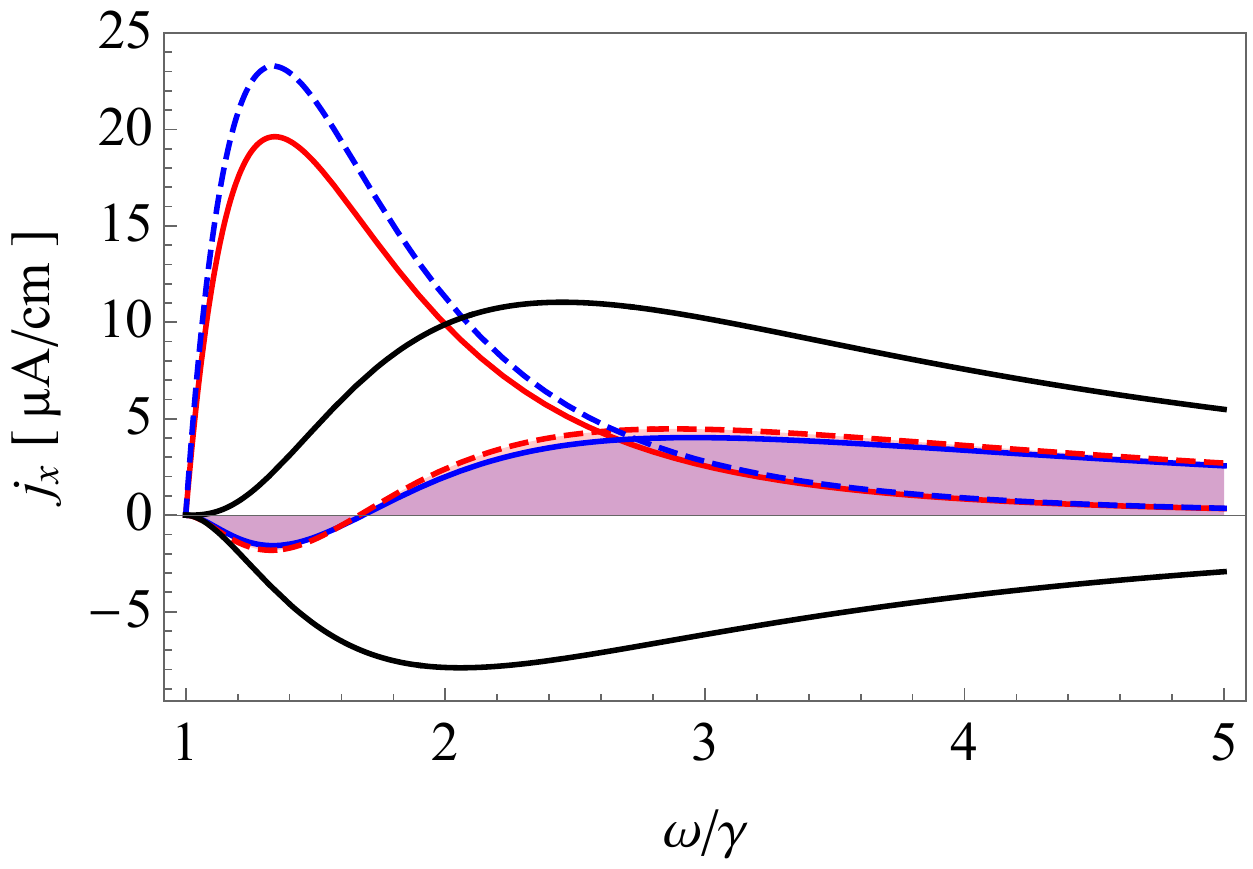}
\caption{Spectrum of electric current density due to the transitions from $m=1$, $\eta=-1$ (solid) and $m=-1$, $\eta=1$ (dashed) impurity states to spin-up conduction band states for the polarizations of light $\sigma = -1$ (red) and $\sigma = 1$ (blue); $\gamma=\epsilon^{i}_{1,1,-1}/\hbar$.
The other parameters are taken the same as in Fig.~\ref{Fig2}. }
\label{Fig3}
\end{figure}

Considering  the $m = 0$ impurity state, we find the corresponding matrix element,
\begin{eqnarray}
\label{EqM0}
|M_0(\textbf{p,k})|^2 &=&
(evA_0)^2
(\hbar v)^2
\frac
{2\pi\epsilon^i_{s,0,\eta}}
{\Delta}
\\
\nonumber
&&
\times
\left\{
\frac{(\eta+\sigma)^2v^2(\textbf{p}-\textbf{k})^2\cos^2\left(\frac{\theta_{s,\eta}}{2}\right)
}{\Big[v^2(\textbf{p}-\textbf{k})^2+\epsilon^i_{s,0,\eta}(\Delta-s\lambda\eta)\Big]^2}
\right.
\\
\nonumber
&&~~~~~~~\left.
+
\frac{
(\eta-\sigma)^2(\Delta-s\lambda\eta)^2\sin^2\left(\frac{\theta_{s,\eta}}{2}\right)
}{\Big[v^2(\textbf{p}-\textbf{k})^2+\epsilon^i_{s,0,\eta}(\Delta-s\lambda\eta)\Big]^2}
\right\}.
\end{eqnarray}
Without the loss of generality, let us choose $\mathbf{k}$ to be directed along the x-axis ($\mathbf{k}=k\hat{x}$). 
Then, in Eq.~\eqref{EqCurrentMain} only the term containing $\cos\varphi_\textbf{p}$ survives and $j_y = 0$, that reflects the fact that the photon-drag current should be directed along the photon wave vector. 
Substituting Eq.~\eqref{EqM0} in Eq.~\eqref{EqCurrentMain}, we find
\begin{equation}
\begin{split}
&
j_{0x}
=
\beta_{0}'\Theta[\delta\hbar \omega_{s,0,\eta}]
\frac{k\pi}{v}\frac{(\Delta-s\lambda\eta)+\delta\hbar \omega_{s,0,\eta}}{\Big[(\delta\hbar \omega_{s,0,\eta})^2 + (\Delta - s\lambda\eta)\hbar \omega\Big]^2}
\\
&
\times
\frac{\delta\hbar \omega_{s,0,\eta}}{(\Delta - s\lambda\eta) + \delta\hbar \omega_{s,0,\eta}}
 \Bigg\{
\frac{4(\Delta-s\lambda\eta)^2}{(\delta\hbar \omega_{s,0,\eta})^2 + (\Delta - s\lambda\eta)\hbar \omega}
\\
&
\times
\delta\hbar \omega_{s,0,\eta}(\eta-\sigma)^2
+
\Big[(\Delta-s\lambda\eta)+\delta\hbar \omega_{s,0,\eta}\Big](\eta+\sigma)^2
\\
\label{EqCurrent0}
&
\times
\left[
\frac{4\Big((\Delta-s\lambda\eta)+\delta\hbar \omega_{s,0,\eta}\Big)\delta\hbar \omega_{s,0,\eta}}{(\delta\hbar \omega_{s,0,\eta})^2 +(\Delta-s\lambda\eta)\hbar \omega}
-2\right]
\Bigg\},
\end{split}
\end{equation}
where 
$\beta_{0}'=en_i\tau v^2\epsilon_i(evA_0)^2/\hbar\Delta$ and $\delta\hbar\omega_{s,m,\eta}=\hbar\omega-\epsilon^i_{s,m,\eta}$.
Figure~\ref{Fig2} shows the spectrum of electric current density for different polarizations of light and $m=0$, $\eta=-1$.
Interesting to note, that in the case of the polarization of light $\sigma = -1$, the electric current flows in the opposite direction in some region of frequencies and then it changes its direction. 
A similar behavior (of the inversion of the direction of the electric current density) was demonstrated in work~\cite{PhysRevB.81.165441}.
Mathematically, it happens due to an interplay of different terms in Eq.~\eqref{EqCurrent0}, shown as dashed curves.
Such behavior is not observed for $\sigma=1$. For the case $m = 1$ (and, correspondingly, $\eta=-1$), we find
\begin{eqnarray}
\label{EqM1}
|M_{1}(\textbf{p},\textbf{k})|^2&=&
(evA_0)^2
(\hbar v)^2
\frac
{2\pi(\Delta^2-\lambda^2)}
{\Delta\epsilon^i_{s,1,-1}}
\\
\nonumber
&\times&
\left\{
\frac{(\sigma-1)^2(\epsilon^i_{s,1,-1})^2\cos^2\left(\frac{\theta_{s,-1}}{2}\right)}{\Big[v^2(\textbf{p}-\textbf{k})^2+\epsilon^i_{s,1,-1}(\Delta+s\lambda)\Big]^2}\right.\\
\nonumber
&&~~~~~\left.+
\frac{(\sigma+1)^2v^2(\textbf{p}-\textbf{k})^2\sin^2\left(\frac{\theta_{s,-1}}{2}\right)}{\Big[v^2(\textbf{p}-\textbf{k})^2+\epsilon^i_{s,1,-1}(\Delta+s\lambda)\Big]^2}
\right\}.
\end{eqnarray}
%
%
Again, only the x-component of the current is finite,
\begin{equation}
\label{EqCurrent1}
\begin{split}
j_{1x}
=
\beta_1'\Theta[\delta\hbar\omega_{s,1,-1}]\frac{k\pi}{v}\frac{(\Delta+s\lambda)+\delta\hbar\omega_{s,1,-1}}{\Big((\delta\hbar\omega_{s,1,-1})^2 + (\Delta + s\lambda)\hbar \omega\Big)^2}
\\
\times
\frac{\delta\hbar\omega_{s,1,-1}}{(\Delta+s\lambda) + 2\delta\hbar\omega_{s,1,-1}}
 \Bigg\{
\frac{(\Delta+s\lambda)+\delta\hbar\omega_{s,1,-1}}{(\delta\hbar\omega_{s,1,-1})^2 + (\Delta + s\lambda)\hbar \omega}
\\
\times
4(\epsilon{^i_{s,1,-1}})^2(\sigma-1)^2+\delta\hbar\omega_{s,1,-1}(\sigma+1)^2
\\
\times
\left(
\frac{4\Big((\Delta+s\lambda)+\delta\hbar\omega_{s,1,-1}\Big)\delta\hbar\omega_{s,1,-1}}{(\delta\hbar\omega_{s,1,-1})^2 +(\Delta+s\lambda)\hbar \omega}
-2\right)
 \Bigg\},
\end{split}
\end{equation}
where $\beta_{1}'=\beta_{0}'[\Delta^2-\lambda^2]/\epsilon_i^2$. 

Figure~\ref{Fig3} shows the spectrum of electric current density for $m = -1$ and $m = 1$. 
We choose $\eta = -1$ for $m = 1$ and $\eta = 1$ for $m = -1$ .
Also, it is the $\sigma=1$ polarization which gives the region of positive and negative electric currents for $m = 1$ (blue solid curve) and for $m = -1$, the $\sigma = -1$ polarized light gives such electric current (red dashed curve).
For a given $\sigma$, we have optical transitions in both the K (for $m=0$ or 1) and K$'$ (for $m=1$ or 0) valleys. 
They can sum up or partially compensate each other.




\textit{Symmetry analysis of the electric current density.}
Let us now analyse the formulas for the electric current density [Eqs.~\eqref{EqCurrent0} and~\eqref{EqCurrent1}] from the symmetry point of view, and compare them with the case of a graphene monolayer~\cite{glazov2014high}.
A single-layer graphene (without a substrate) possesses the $D_{6h}$ point group, while single-layer MoS$_2$ has $D_{3h}$ point group. 
However, for both the groups, the fourth-rank (generalized conductivity) tensor $\Phi_{\alpha\beta\gamma\mu}$ is the same~\cite{boyd2020nonlinear}.

The general expression for the electric current density reads~\cite{glazov2014high}
\begin{equation}
\label{EqCur1x}
j_{x} = T_1 k_x \frac{|E_x|^2 + |E_y|^2}{2} +
T_2 k_x \frac{|E_x|^2 - |E_y|^2}{2},
\end{equation}
where $E_x$ and $E_y$ are the components of the electric field; 
$T_1$ and $T_2$ are constants describing linear photon drag effect. 
Since the electric field is circularly polarized in our case, $\mathbf{E}=E_0(1,i)$,
only the first term in Eq.~\eqref{EqCur1x} remains, and the other one vanishes since $|E_x|=|E_y|$. 
We see that Eqs.~\eqref{EqCurrent0} and~\eqref{EqCurrent1} obey the symmetry properties of the system.


\textit{Light absorption coefficient. }
Furthermore, let us study the light absorption coefficient for the $m-$th impurity state.
It is defined as the ratio of the energy flux of absorbed photons and the average energy flux of incident photons~\cite{fang2013quantum}, $\alpha_m(\hbar \omega) = \hbar \omega W_m/P$,
%
%
where $P$ is the average of the Poynting flux for the light intensity~\cite{chuang2012physics}, $P=n_rc\epsilon_0\omega^2A_0^2/2$, 
%
%
where $n_r$ is the refractive index of MoS$_2$ and $\epsilon_0$ is the vacuum permittivity.
The probability of light absorption in a given valley $\eta$ and from a particular impurity state $m$ is given by the Fermi golden rule,
\begin{equation}
\nonumber
W_m(\omega)=\frac{2\pi n_i}{\hbar}\int\frac{d\textbf{p}}{(2\pi\hbar)^2}|M_m(\textbf{p},0)|^2\delta(E_{c}(\textbf{p})-E_i-\hbar \omega).
\end{equation}
For the transition from $m=0$ impurity state, we find
\begin{equation}
\begin{split}
&\alpha_0 = 
\frac{2\pi n_ie^2v^2\epsilon^i_{s,0,\eta}}{n_rc\omega\Delta\epsilon_0}
\Theta[\delta\hbar \omega_{s,0,\eta}]
\frac{\delta\hbar \omega_{s,0,\eta}}{(\Delta - s\lambda\eta)+2\delta\hbar \omega_{s,0,\eta}}
~~~
\\
&\times
\frac{\sqrt{4\Big(\Delta-s\lambda\eta+\delta\hbar \omega_{s,0,\eta}\Big)\delta\hbar \omega_{s,0,\eta}+(\Delta-s\lambda\eta)^2}}{\Big((\delta\hbar \omega_{s,0,\eta})
 ^2+(\Delta-s\lambda\eta-\epsilon^i_{s,0,\eta})\hbar \omega\Big)^2}
\\
&\times
\left\{
(\eta+\sigma)^2\Big(\Delta-s\lambda\eta+\delta\hbar \omega_{s,0,\eta}\Big)^2\right.
\\
\nonumber
&~~~~~~~~~~~~~~~~~~~~~\left.
 +(\eta-\sigma)^2(\Delta - s\lambda\eta)^2
 \right\},
\end{split}
\end{equation}
and for $m=1,\eta=-1$ state,
\begin{equation}
\begin{split}
\alpha_1 & = 
\frac{2\pi n_ie^2v^2}{n_rcw\Delta\epsilon_0}
\Theta[\delta\hbar \omega_{s,1,-1}]
\frac{\Delta^2-\lambda^2}{\epsilon^i_{s,1,-1}}
%
%
\frac{(\Delta+s\lambda)+\delta\hbar \omega_{s,1,-1}}{(\Delta + s\lambda)+2\delta\hbar \omega_{s,1,-1}}
\\\times&
\frac{\sqrt{4\Big(\Delta+s\lambda+\delta\hbar \omega_{s,1,-1}\Big)\delta\hbar \omega_{s,1,-1}+(\Delta+s\lambda)^2}}{\Big((\delta\hbar \omega_{s,1,-1})
 ^2+(\Delta+s\lambda-\epsilon^i_{s,1,-1})\hbar \omega\Big)^2}
\\\times&
\left\{
(\sigma-1)^2(\epsilon^i_{s,1,-1})^2
 +(\sigma+1)^2(\delta\hbar \omega_{s,1,-1})^2
 \right\}.
\end{split}
\end{equation}
\begin{figure}[t!]
\centering
\includegraphics[width=0.4\textwidth]{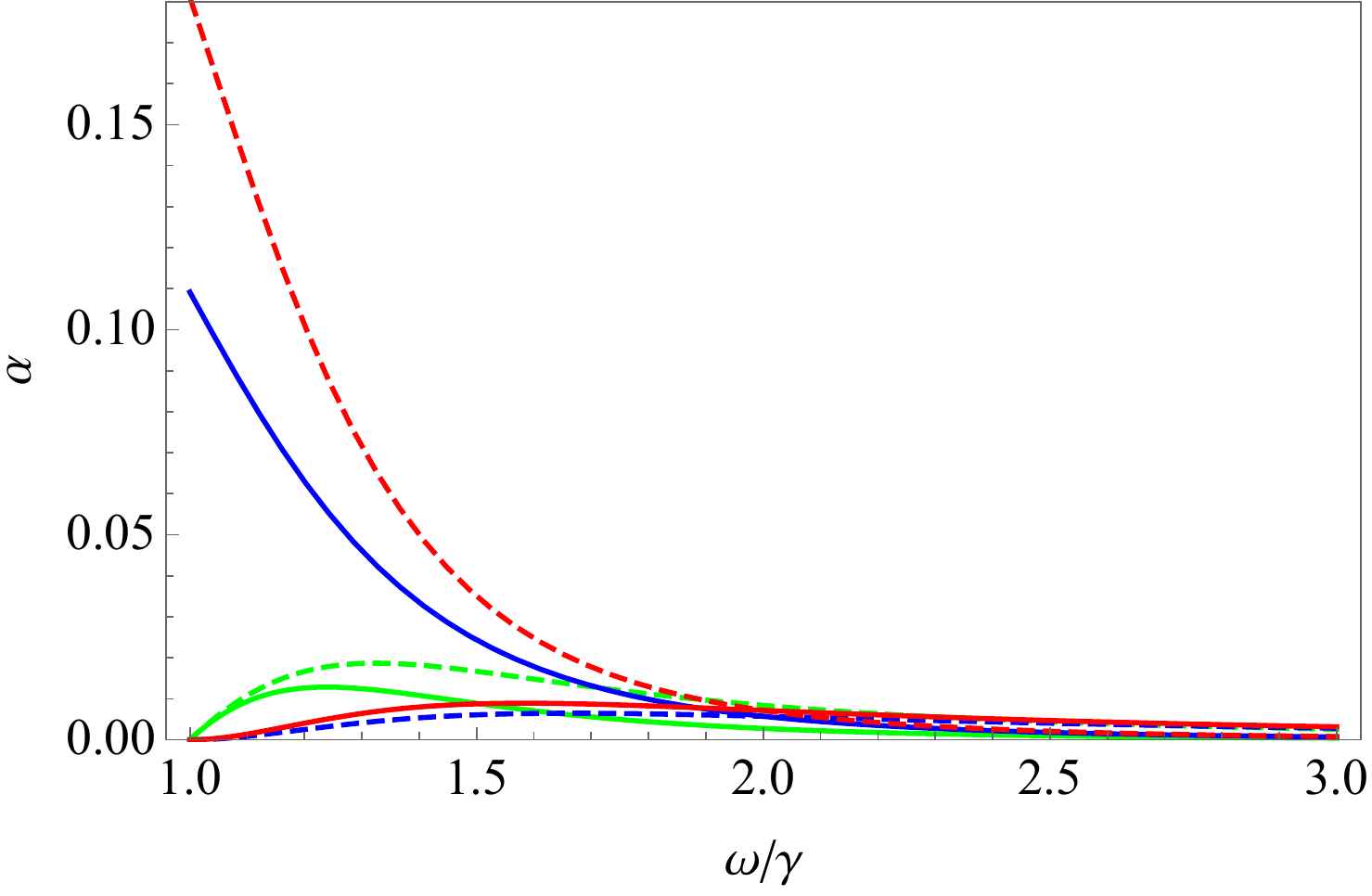}
\caption{Spectrum of
absorbance for $m = -1$ (red), $m = 0$ (green), and $m = 1$ (blue); $\sigma = 1$ (dashed) and $\sigma = -1$ (solid).
}
\label{Fig4}
\end{figure}
%
%

Figure~\ref{Fig4} shows the spectra of absorbance. 
For the transitions from the state $m = 1$, the magnitude of absorbance is higher for the $\sigma = -1$ light but by increasing the photon energy, 
the valley dependence disappears.
For the transitions from the state $m = 0$, both the polarized lights give comparable contribution.
%

It is enlightening to compare the matrix element corresponding to impurity-band transitions with the matrix element for the interband transitions, $|M_{cv}(\mathbf{}{p})|^2$~\cite{Kovalev_2018, PhysRevB.103.035434}. 
The valley selectivity  for interband transitions is to a large extent satisfied only at small values of momentum $p$, giving $|M_{cv}(0)|^2\propto(\eta+\sigma)^2$. 
In our case, the transitions from $m=0$ impurity states are strongly suppressed due to $|M_0(0,0)|^2\rightarrow0$, whereas for $m=\pm1$ we find $|M_{m=\pm1}(0,0)|^2\propto\epsilon_i^2(\sigma+\eta)^2$ under the condition $m+\eta=0$. 
It means that the valley selectivity takes place for orbital impurity states $m=\pm1$ (and thus, we have $\exp(im\varphi)\neq1$), reflecting the chirality of the band electron wavefunction. These general conclusions are supported by the numerical analysis. 
For instance, the absorption coefficient in Fig.~\ref{Fig4} is large in the vicinity of the threshold for $m=1,\eta=-1$ state at $\sigma=-1$ polarization.
%
%
%
%
%
%
%
%
%

\textit{
In conclusion}, we 
have studied the selection rules for the light-induced transitions from impurity states to the conduction band in two-dimensional gapped Dirac materials.
For that, we calculated and investigated
the absorption coefficient 
and the photon-drag-induced electric current.
For clarity, we used the shallow impurity potential model.  
Nevertheless, this model correctly reflects the selection rules of any impurity possessing the azimuthal symmetry. 
Thus, our conclusions on the optical selection rules are sufficiently general. 

\textit{Acknowledgements.}
We thank Dr.~Meng Sun and Ihor Vakulchyk for useful discussions.
The part of the calculations devoted to the absorption coefficient was supported by the Russian Science Foundation (Project No.~17-12-01039).
The current density analysis was supported by the Institute for Basic Science in Korea (Project No.~IBS-R024-D1).


\bibliography{main}
\bibliographystyle{apsrev4-1}

\end{document}